# Superconductivity in High-Entropy-Alloy Telluride AgInSnPbBiTe$_5$

Yoshikazu Mizuguchi[1]*

[1] *1-1, Minami-osawa, Hachioji, Tokyo 192-0397, Japan*

A polycrystalline sample of the high-entropy-alloy-type telluride AgInSnPbBiTe$_5$ was synthesized using high-pressure synthesis. Superconductivity with a transition temperature ($T_c$) of 2.6 K was observed in AgInSnPbBiTe$_5$. Elemental and structural analyses revealed that five metals are mixed in a metal site of an NaCl-type structure. Since AgInSnPbBiTe$_5$ has a cation site (Ag, In, Sn, Pb, and Bi) and an anion site (Te), this is the first example of a high-entropy-alloy (HEA) pseudo-binary superconductor.

## 1. Introduction

High-entropy alloys (HEAs), typically defined as alloys containing five or more elements with concentrations between 5 at.% and 35 at.%, have been extensively studied in the field of materials science and engineering.[1,2] For example, structural materials for utilization at high temperatures can be stabilized using high mixing entropy. A HEA superconductor Ta$_{34}$Nb$_{33}$Hf$_8$Zr$_{14}$Ti$_{11}$ with a transition temperature ($T_c$) of 7.3 K has been discovered.[3] Although the observed superconductivity in most HEA superconductors is believed to be conventional and phonon-mediated, interest in HEA superconductors has been growing due to their notable features and possibilities for material development.[4-9] For example, measurements revealed that the superconductivity states in a HEA superconductor are robust at pressures under 190 GPa.[10] This suggests that HEA superconductors could be useful under extreme conditions.

Recently, we have reported on the synthesis and the superconducting properties of REO$_{0.5}$F$_{0.5}$BiS$_2$ with a HEA-type rare earth oxide (REO) blocking layer; the REO$_{0.5}$F$_{0.5}$BiS$_2$ is a typical BiS$_2$-based layered superconductor[11-14]. The rare-earth (RE) site of the samples examined in the study were occupied by five different RE elements. Interestingly, the increase in mixing entropy at the RE site affected the local structure of the conducting layer (BiS$_2$ layer). In the BiS$_2$-based compounds, the suppression of local disorder in the BiS$_2$ layers, whose disorder is most likely due to the presence of a Bi-6p lone pair in the conducting layer, is essential to induce bulk superconductivity.[15-17] Notably, the local disorder is suppressed by the increase in mixing entropy at the RE site.[18] Thus, further investigation into the influence of HEA states on crystal structures, superconducting properties, and other functionalities is needed for materials with a complicated structure, such as those with a layered structure or intermetallic compounds.

In this study, we focus on a simpler NaCl-type (#225, $Fm$-$3m$, $O_h^5$) telluride to explore new HEA-type superconductors because several metal tellurides have a NaCl-type structure. Since the structure of an NaCl-type telluride is composed of both a cationic metal site and an anionic tellurium site, a HEA-type telluride is a suitable material to use to investigate the influence of HEA states on the states of a tellurium site. The knowledge obtained from studying HEA tellurides will be useful for further developing HEA-type materials. In this paper, we report on the synthesis of AgInSnPbBiTe$_5$ with an NaCl-type structure and the observation of a superconducting transition with a $T_c$ of 2.6 K. Since the metal site in AgInSnPbBiTe$_5$ contains Ag, In, Sn, Pb, and Bi, it can be regarded as a *HEA-type telluride superconductor*.

## 2. Experimental Details

Polycrystalline samples of AgInSnPbBiTe$_5$ were obtained using high-pressure annealing of AgInSnPbBiTe$_5$ precursor powders, which were obtained by melting Ag powders (99.9%) and In (99.99%), Sn (99.999%), Pb (99.9%), Bi (99.999%), and Te (99.999%) grains with a nominal composition of AgInSnPbBiTe$_5$ at 800 °C in an evacuated quartz tube. The high-pressure annealing was performed using a cubic-anvil-type 180-ton press (CT factory) under 3 GPa for 30 min. Three different annealing conditions were examined. The obtained samples are labeled A, B, and C for annealing temperatures of 500, 600, and 700 °C, respectively. The chemical composition of the samples was examined by energy-dispersive X-ray spectroscopy (EDX) on a TM-3030 (Hitachi Hightech) equipped with an EDX SwiftED analyzer (Oxford). The EDX analysis was performed at five different points with a typical analysis area of 30 μm × 30 μm. The standard deviation represents the analysis error of the chemical composition. Powder X-ray diffraction (XRD) was performed using a MiniFlex600 (RIGAKU) with CuKα radiation and a D/teX-Ultra detector and a conventional $\theta$-$2\theta$ method. The resulting XRD pattern was refined using the Rietveld analysis and a RIETAN-FP.[19] The refinement was performed with fixed occupation parameters for both sites as the chemical composition was assumed to be AgInSnPbBiTe$_5$.

To investigate the superconducting properties of AgInSnPbBiTe$_5$, the temperature dependence of magnetic susceptibility, magnetization loop ($M$-$H$ loop), and temperature dependencies of electrical resistivity under magnetic fields were measured. Magnetic susceptibility was measured using a superconducting quantum interference devise (SQUID) with MPMS3 (Quantum Design) after both zero-field cooling (ZFC) and field cooling (FC) occurred. Resistivity was measured on a GM refrigerator system (Made by Axis) using a conventional four-probe method. The Au wires (25 μm in diameter) were attached to the samples' surface using Ag paste.

3. Results and Discussion

The color of the samples was silver. As shown in Fig. S1 (Supplemental Materials), the surface of the samples appears homogeneous, and the typical grain size is larger than 10 μm.[20] The average chemical compositions for sample A, B, and C were $Ag_{0.99(5)}In_{0.99(12)}Sn_{1.10(7)}Pb_{0.98(8)}Bi_{1.01(4)}Te_{4.92(4)}$, $Ag_{1.05(2)}In_{1.04(2)}Sn_{1.12(4)}Pb_{1.01(6)}Bi_{0.80(4)}Te_{4.99(3)}$ and $Ag_{1.04(3)}In_{1.02(5)}Sn_{1.04(3)}Pb_{0.88(5)}Bi_{1.10(9)}Te_{4.93(5)}$, respectively; these compositions were determined by averaging the EDX analysis result from five points. For all the samples, the composition at the metal site satisfies the HEA criterion of falling in the range of 5 at.% to 35 at.%, and we consider the metal site of $AgInSnPbBiTe_5$ to be in the HEA state. Since the composition of the samples is close to the nominal composition, we call the samples $AgInSnPbBiTe_5$ in this study. Due to the simple NaCl-type structure, there is possibility of mixing of the metals (Ag, In, Sn, Pb, or Bi) and Te. However, from the EDX analysis, we confirmed that the Te to metal ratio is close to 50% for all the investigated points. We found that compositional trend changes depending on the annealing temperature. For sample A, the composition at the metal site is close to the nominal composition, but Bi-poor and Pb-poor compositions were found for sample B and C. The slightly different compositions affect the lattice constant ($a$) as displayed in the next paragraph.

Figure 1(a) shows the XRD patterns for all the samples. The XRD peaks of the major phase can be indexed by the NaCl-type structural model. The XRD profile for sample A suggests a single phase. However, as shown in Fig. 1(b), we found that the peaks for sample B and C have an anisotropic structure with a shoulder at a lower angle. This is due to the presence of a minor phase with a larger lattice constant. This assumption could be proofed by the improvement of Rietveld refinement for sample C by using two-phase analysis (Fig. S2 of the Supplemental Materials)[20]. The two-phase Rietveld refinement for sample C [$a$ = 6.2426(2) Å] with a minor cubic phase with $a$ = 6.35 Å resulted in the lowest reliability factor $R_{wp}$ of 8.5%, while single-phase analysis resulted in $R_{wp}$ = 10.5%.

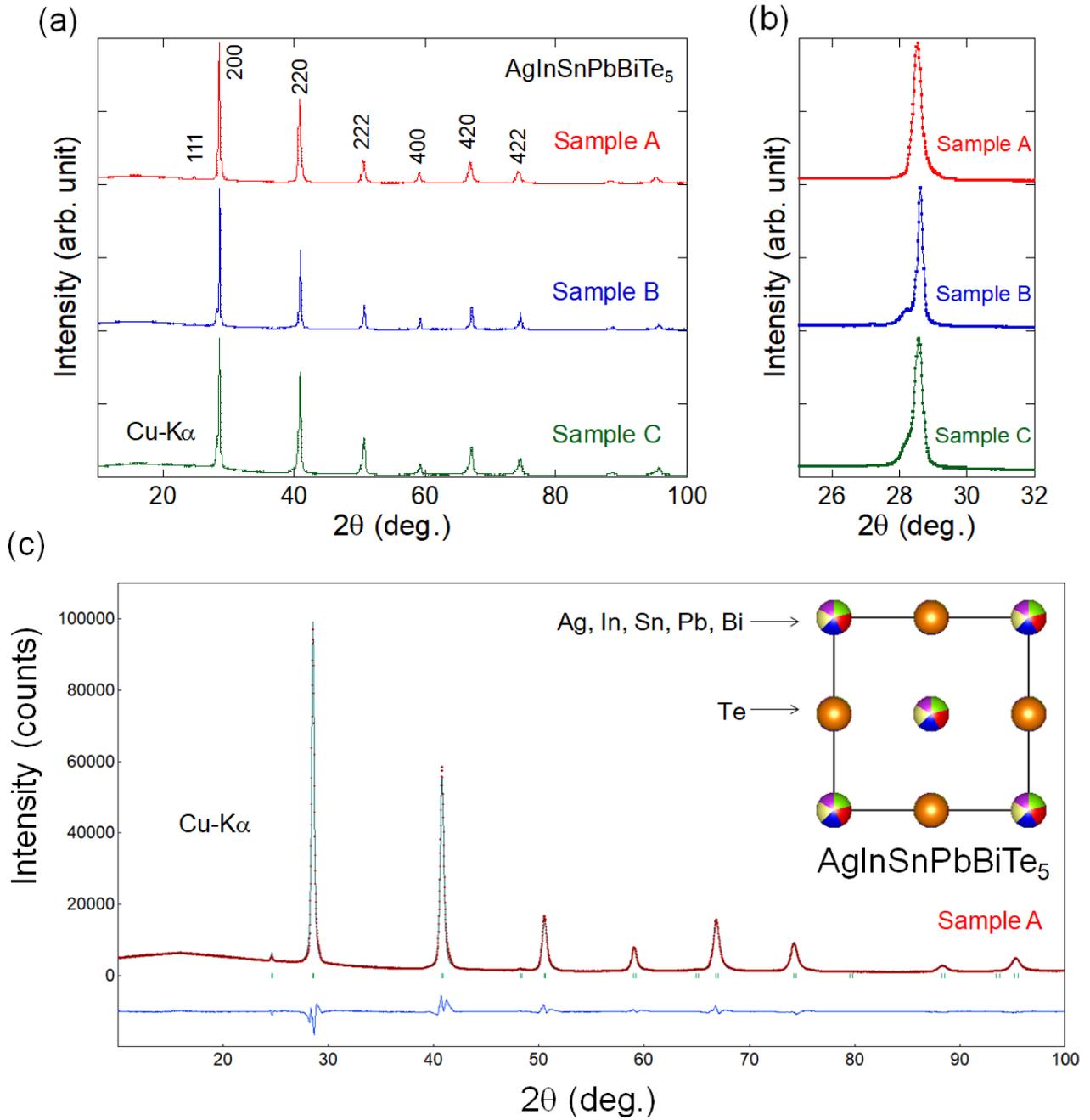

Fig. 1. (color online) (a) XRD patterns for sample A, B, and C with Miller indices. (b) Zoomed profiles near 200 peak. (c) Rietveld refinement fitting of the XRD pattern for sample A. The inset shows the schematic image of the crystal structure of AgInSnPbBiTe$_5$, which was drawn using the refinement results and VESTA software.[21]

**Table I. Structural parameters for AgInSnPbBiTe$_5$ (sample A) determined from Rietveld refinement.**

| Phase | AgInSnPbBiTe$_5$ |
|---|---|
| Space group | $Fm$-$3m$ (#225) |
| Lattice constant | 6.25546(11) Å |
| Atomic coordinate (Ag, In, Sn, Pb, and Bi) | (0, 0, 0) |
| Atomic coordinate (Te) | (0.5, 0.5, 0.5) |
| $U_{iso}$ (Ag, In, Sn, Pb, and Bi) | 0.0129(9) Å$^2$ |
| $U_{iso}$ (Te) | 0.0036(9) Å$^2$ |
| $R_{wp}$ | 6.8% |

Figures 2(a-c) show the temperature dependences of the magnetic susceptibility of AgInSnPbBiTe$_5$ (sample A, B, and C). Diamagnetic signals corresponding to the emergence of superconducting states are observed below 2.5–2.6 K. The estimated shielding fractions estimated from ZFC values at 1.8 K were about 150%, 100%, and 110%, respectively. The shielding fraction largely exceeding 100% is due to the plate-like shape of the samples leading to a demagnetization effect. Based on the large shielding fraction and the observation of the typical $M$-$H$ loop [Fig. 2(d)], we were able to confirm the superconducting transition in the AgInSnPbBiTe$_5$ sample. From the magnetic measurements and the XRD results, we consider sample A is the most homogeneous.

Figure 3(a) shows the temperature dependences of the electrical resistivity of AgInSnPbBiTe$_5$ (sample A and C). The temperature dependence is metallic, but the residual resistivity ratio (RRR) is very small. This behavior is similar to a conventional HEA superconductor.[3] As shown in Fig. 3(b), resistivity begins to decrease at 2.8 K, which corresponds to the onset temperature ($T_c^{onset}$), and is zero at 2.55 K ($T_c^{zero}$) for sample A. In Figs. 4(a,b), temperature dependences of electrical resistivity for sample A and C measured under $H$ = 0, 2000, 4000, and 6000 Oe are shown. $T_c$ decreases with an increase in the magnetic field. To investigate the superconducting properties under the magnetic fields, $T_c$-$\rho$ (90%) was estimated as the temperature where resistivity becomes 90% of normal state resistivity at 4 K [see Figs. 4(a,b)]. Figures 4(c,d) show the magnetic field-temperature phase diagram of AgInSnPbBiTe$_5$. Based on a linear fitting of the data, $H_{c2}$ (0 K) is estimated as 28000 and 24000 Oe for sample A and C, respectively. Furthermore, using the Werthamer-Helfand-Hohemberg (WHH) model., which can be applied to superconductors in a dirty limit, $H_{c2}$ (0 K) ~ 19000 Oe for sample A.[22]

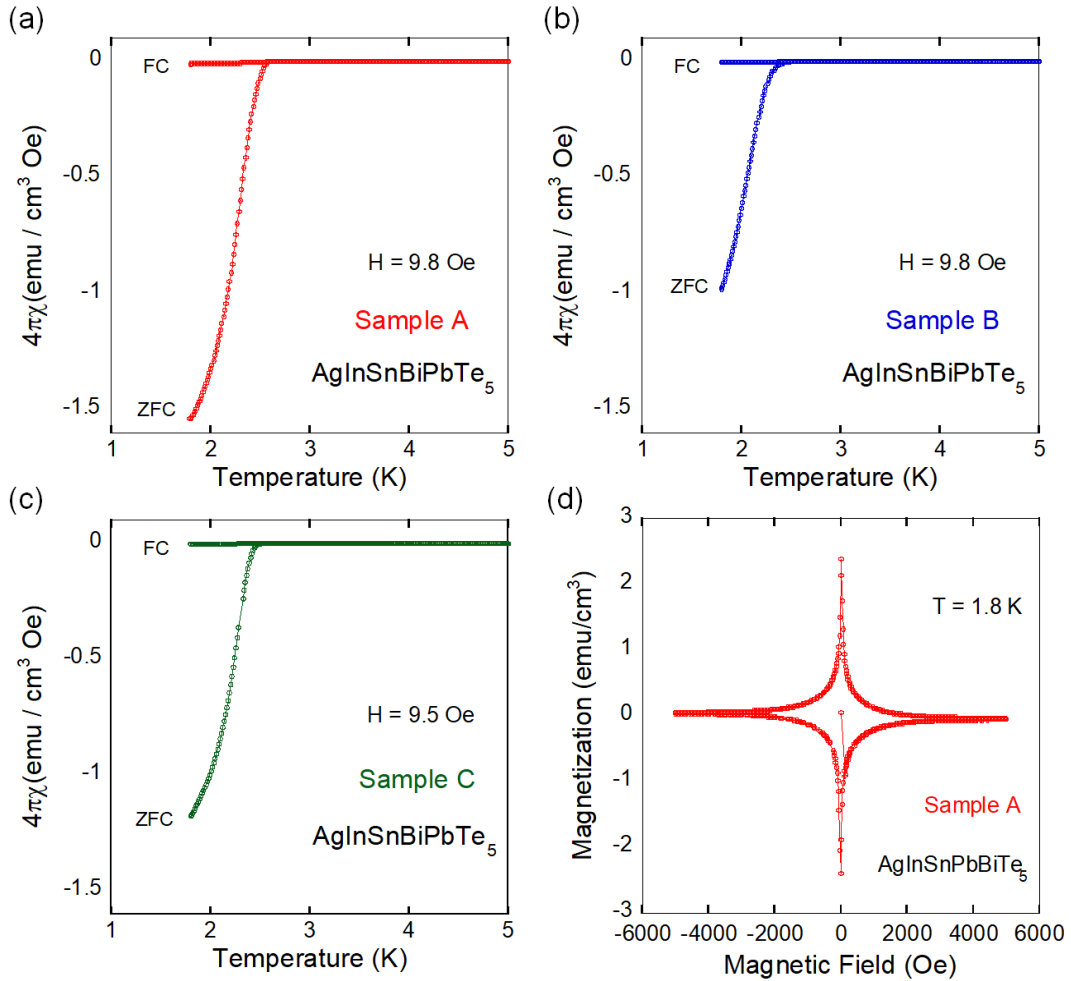

Fig. 2. (color online) (a-c) Temperature dependence of magnetic susceptibility for sample A, B, and C. (d) *M-H* loop at 1.8 K for sample A.

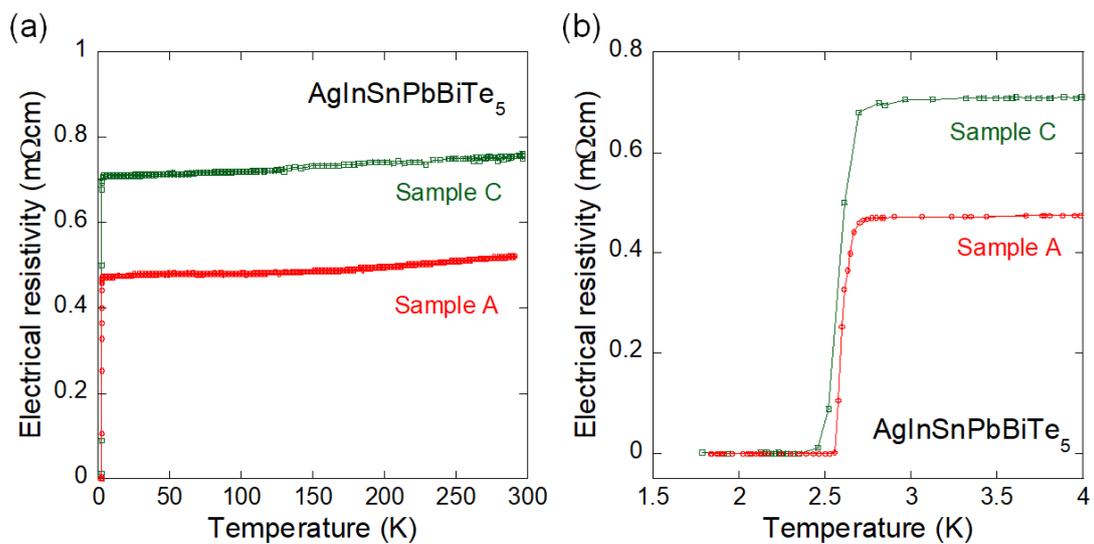

Fig. 3. (color online) (a) Temperature dependence of electrical resistivity for sample A and C. (b) Zoomed in view of (a) near the superconducting transition.

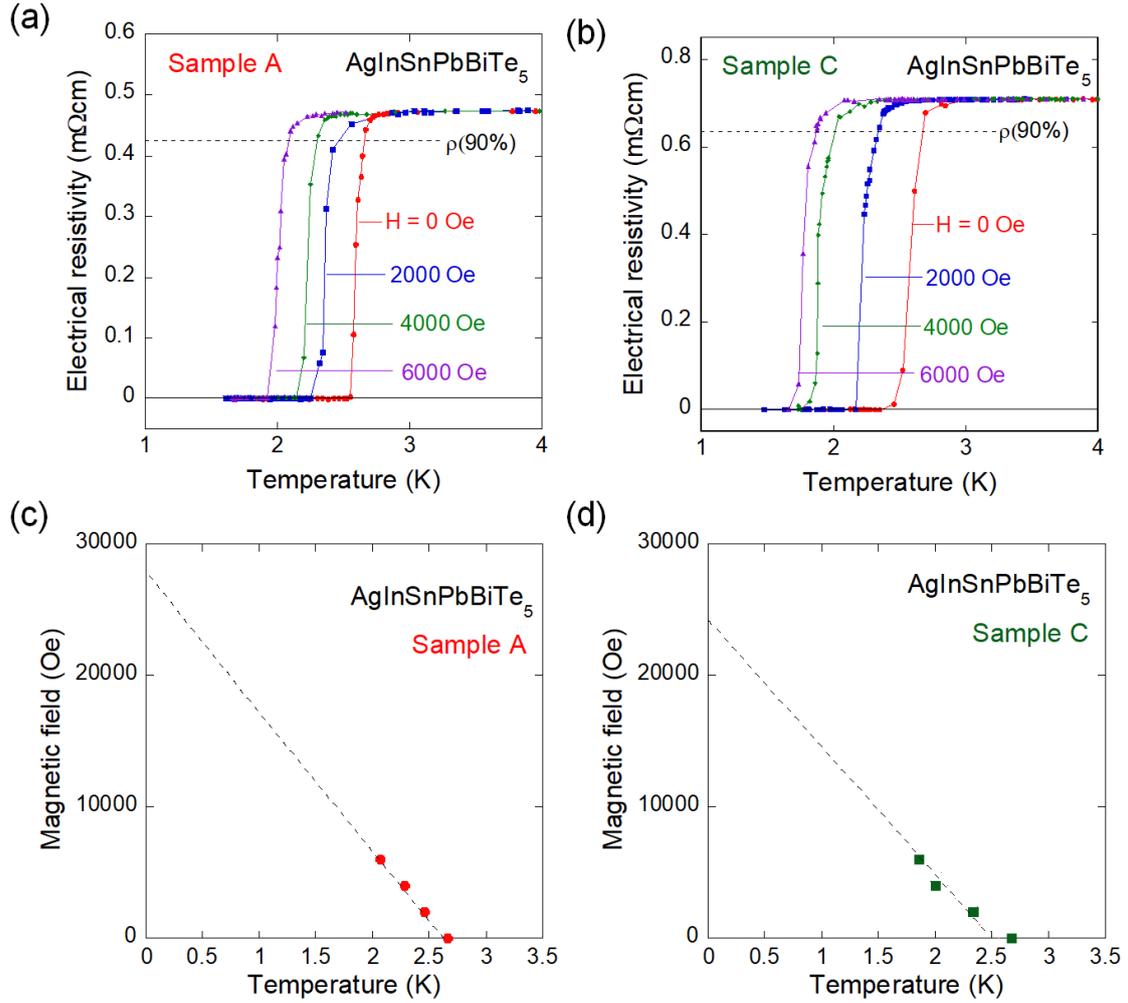

Fig. 4. (color online) (a,b) Temperature dependencies of electrical resistivity under magnetic fields for sample A and C. (c,d) Magnetic field-temperature phase diagram for sample A and C.

The impact of the discovery of superconductivity in a HEA-type telluride is an important point of discussion. Thus far, many HEA superconductors have been discovered in cubic structures (BCC, HCP, and CsCl-type).[4] However, these HEA superconductors contain metal elements only. The AgInSnPbBiTe$_5$ has a NaCl-type structure, which is composed of a cation site (Ag, In, Sn, Pb, and Bi) and an anion site (Te). Therefore, we can expect quite different chemical and electronic states between these sites. On the metal site characteristics, Pb and Sn prefer a valence state of +2 in an NaCl-type structure.[23,24] In addition, AgBiCh$_2$ can be crystalized in a NaCl-type structure at high temperatures by alloying Ag$^+$ and Bi$^{3+}$.[25,26]. Furthermore, InTe can be crystalized in an NaCl-type structure by applying the pressure effect.[24,27] In an NaCl-type structure, the In valence state is the mixed valence of In$^+$ and In$^{3+}$. Therefore, the present composition satisfies electronic neutrality because the average valence of the metal site can be considered to be +2. This assumption leads to the valence states of o$M^{2+}$Te$^{2-}$ where $M$ is Ag, In, Sn, Pb, and Bi. In this respect, to our best knowledge, AgInSnPbBiTe$_5$ is the first example of a HEA-type pseudo-binary superconductor that is not a simple alloy.

Because of the expected exotic superconducting properties of the HEA-type telluride AgInSnPbBiTe$_5$, there is a possibility for it to be an extremely disordered superconductor like AgSnSe$_2$, which has a mixed valence Sn alloyed with Ag in a single site.[28-30] Using the material design concept of the HEA telluride, we can tune the disorder by changing the number of elements in the metal site. Such an investigation may be useful for understanding the relation between valence-skipping states, electronic/structural local inhomogeneity, and superconductivity mechanisms, which have been discussed for tellurides like InTe and Pb$_{1-x}$Tl$_x$Te.[24,27,31]

## 4. Summary

We have synthesized new HEA-type superconductors, AgInSnPbBiTe$_5$, using high-pressure synthesis. AgInSnPbBiTe$_5$ is crystalized in a cubic NaCl-type structure. Since the cationic metal site is occupied with five different elements with a similar occupancy (about 20%), this material can be regarded as a HEA-type telluride. To the best of our knowledge, AgInSnPbBiTe$_5$ is the first example of a HEA-type pseudo-binary compound that is not a simple alloy. Based on magnetic susceptibility and electrical resistivity measurements, a superconducting transition with a $T_c$ of 2.6 K was confirmed for AgInSnPbBiTe$_5$. This study provides the first example of a HEA-type pseudo-binary superconductor. Although the observed $T_c$ is very low (2.6 K), we expect a higher $T_c$ or unconventional properties of superconductivity in another compound in the HEA telluride family. To ascertain a strategy to increase the $T_c$ in HEA materials, an investigation into the local structure of both the metal and Te sites is needed. These findings are useful for further development of various HEA-type compounds including HEA-type layered compounds.

**Acknowledgments**

The author would like to thank O. Miura, M. Katsuno, and R. Jha for their experimental support. This work was partly supported by the Grants-in-Aid for Scientific Research by JSPS (Nos. 15H05886 and 18KK0076) and the Advanced Research Program under the Human Resources Funds of Tokyo.

**(Supplemental Material)**
# Superconductivity in High-Entropy-Alloy Telluride AgInSnPbBiTe$_5$

Yoshikazu Mizuguchi[1]*

[1]*1-1, Minami-osawa, Hachioji, Tokyo 192-0397, Japan*

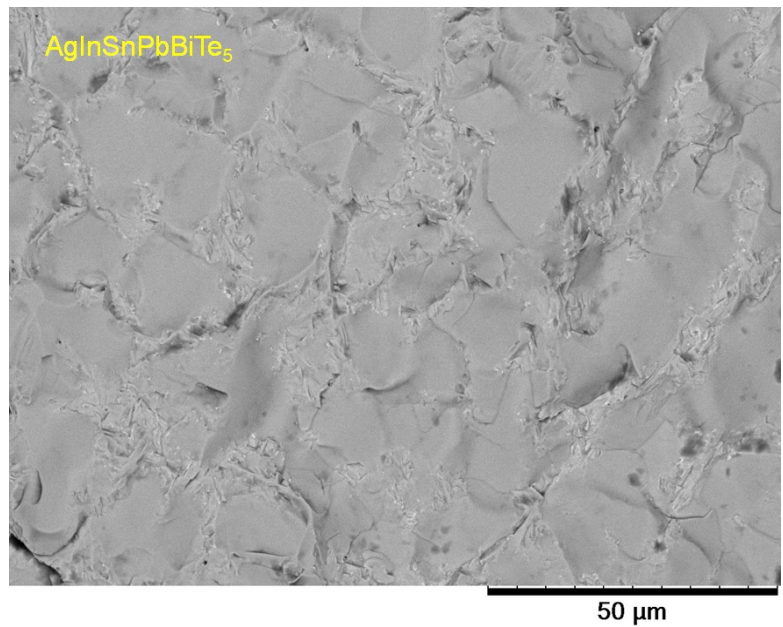

Fig. S1. SEM image for AgInSnPbBiTe$_5$ (sample C).

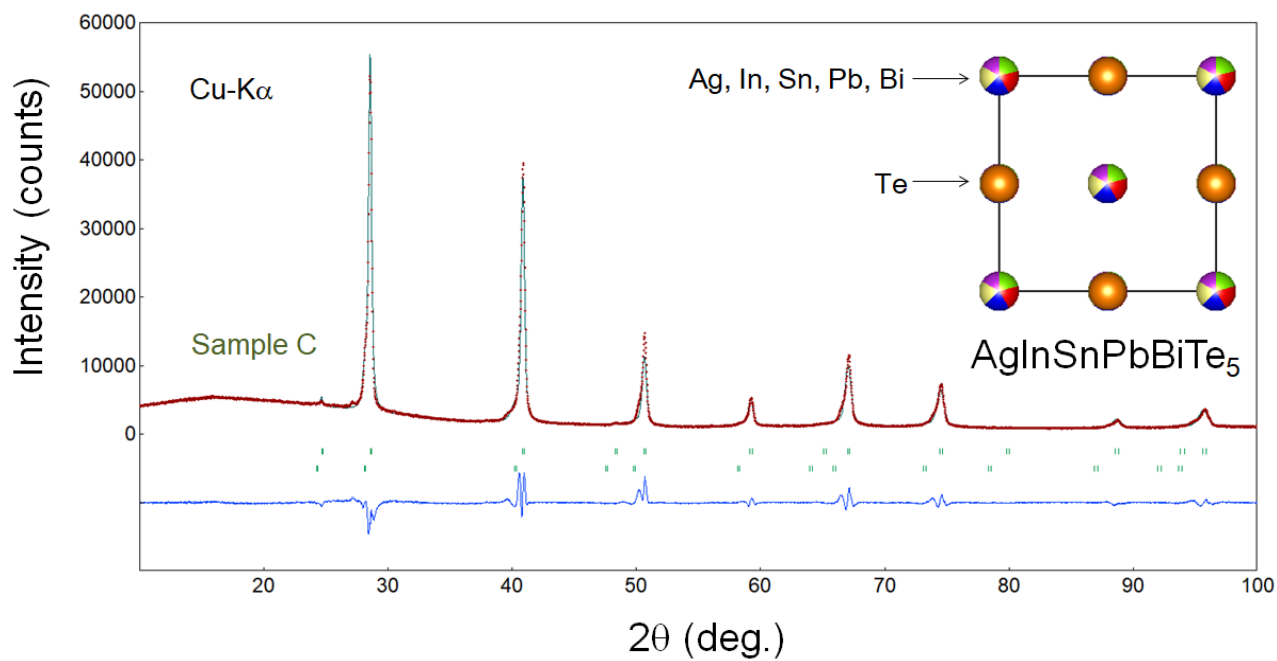

Fig. S2. Two-phase Rietveld refinement result for AgInSnPbBiTe$_5$ (sample C). A minor phase with a cubic (NaCl-type) structure with $a$ = 6.35 Å was assumed.

11